 \documentclass[final,1p,times]{elsarticle}
 \usepackage{graphics}
\usepackage{graphicx}
 \usepackage{epsfig}
\usepackage{textcomp}
\usepackage{booktabs}
\usepackage{amssymb}
\begin{document}

\begin{frontmatter}



\title{The crystal and magnetic structure of the magnetocaloric compound FeMnP$_{0.5}$Si$_{0.5}$}


%\author{author list}

\author[mkem]{Viktor H\"{o}glin\corref{cor1}}\ead{viktor.hoglin@mkem.uu.se}
\author[ftf]{Matthias Hudl}
\author[mkem]{Martin Sahlberg}
\author[ftf]{ Per Nordblad}
\author[NPI]{Premysl Beran}
\author[mkem]{Yvonne Andersson}

\cortext[cor1]{Corresponding author.}

%\address{adress list}
\address[mkem]{Department of Materials Chemistry, Uppsala University, Box 538, 75121 Uppsala, Sweden}
\address[ftf]{Department of Engineering Sciences, Uppsala University, Box 534, 75121 Uppsala, Sweden}
\address[NPI]{Nuclear Physics Institute, Academy of Sciences of the Czech Republic, 25068 Rez, Czech Republic}

\begin{abstract}
The crystal and magnetic structure of the magnetocaloric compound FeMnP$_{0.5}$Si$_{0.5}$ have been studied by means of neutron and X-ray powder diffraction. Single phase samples of nominal composition FeMnP$_{0.5}$Si$_{0.5}$ have been prepared by the drop synthesis method. The compound crystallizes in the Fe$_2$P-type structure (P\=62m) with the magnetic moments aligned along the a-axis. It is found that the Fe atoms are mainly situated in the tetrahedral 3g site while the Mn atoms prefer the pyramidal 3f position. The material is ferromagnetic (T$_C$ = 382 K) and at 296 K the total magnetic moment is 4.4 $\mu_B$/f.u. It is shown that the magnetic moment in the 3f site is larger (2.5 $\mu_B$) than in the 3g site (1.9 $\mu_B$).

\end{abstract}


\end{frontmatter}

% \linenumbers


\section{Introduction}
\label{}
Magnetocaloric compounds have gained an increased interest since the middle 1990's due to environmental and energy benefits from magnetic refrigeration and the discovery of the Giant Magnetocaloric Effect (GMCE) \cite{gme_1}. In GMCE compounds rare earth metals are common alloying elements and due to their rareness in nature and high costs, they will be a problem in a future large-scale production. Hence, there is a need to find more common and cheaper compounds not based on rare earth metals that possess the GMCE.

A large number of studies on the compound Fe$_2$P have been performed since the 1960's and its crystallographic and magnetic properties have been well investigated \cite{rundqvist_59, wappling2, lundgren_fe2p}.  The relatively large saturation magnetization, first order nature of the transition and readily tunable transition temperature with various substitutions make the Fe$_2$P system a candidate compound for magnetocaloric applications. Numerous compounds based on Fe$_2$P have been fabricated during the years and compounds of the FeMnP$_{1-x}$M$_x$-type (M = Si, Ge and/or As) have shown improved  magnetocaloric properties.

In this investigation, samples of nominal composition FeMnP$_{0.5}$Si$_{0.5}$ have been synthesized and the crystal and magnetic structure and the magnitude of the magnetic moments at different temperatures have been determined.  There is a controversy about the exact properties of the FeMnP$_{1-x}$Si$_x$-system.  A previous study by Cam Thanh et al. \cite{cam_tanh_1} on samples of nominally the same composition, FeMnP$_{0.5}$Si$_{0.5}$, prepared by a ball milling technique, had a transition temperature of 332 K with higher preserved magnetocaloric effect.   There is a significant difference in the structural and magnetic properties presented by Cam Thanh et al. \cite{cam_tanh_1} compared to our results.  This dichotomy may be explained by the fact that their samples contained about 18\% of a second phase, identified as Fe$_{2}$MnSi.

Our magnetization measurements on the FeMnP$_{0.5}$Si$_{0.5}$  sample show a ferromagnetic transition temperature of 382 K and the magnetic entropy change $-\Delta S_M$ has from magnetization experiments been estimated to about 8 J/kgK in a magnetic field change of 1.8 T \cite{mossbauer_uu}. The magnetocaloric effect is slightly lower compared to similar compounds \cite{bruck_fe2p_comp, thanh_mnfepsige} but due to its content of only common, non-toxic, elements FeMnP$_{0.5}$Si$_{0.5}$ is a promising magnetocaloric compound within the  FeMnP$_{1-x}$M$_x$ alloy system.

\section{Experiments}
\label{}
\subsection{Sample preparation}
\label{}
FeMnP$_{0.5}$Si$_{0.5}$ samples were prepared by the drop synthesis method \cite{hf_ugn} using a high frequency induction furnace at 1623-1673 K in an Ar atmosphere of 40 kPa. Stoichiometric amounts of iron (Leico Industries, purity 99.995\%. Surface oxides were reduced in H$_2$-gas.),  manganese (Institute of Physics, Polish Academy of Sciences, purity 99.999\%),  phosphorus (Cerac, purity 99.999\%) and silicon (Highways International, purity 99.999\%) were used as raw materials. All samples were crushed, pressed into pellets and sealed in evacuated fused silica tubes. Subsequently, the samples were sintered at 1373 K for 1 h, annealed at 1073 K for 65 h and finally quenched in cold water.

\subsection{X-ray powder diffraction}
Phase analysis and crystal structure characterizations were performed using X-ray powder diffraction (XRD) with a Bruker D8 diffractometer equipped with a V\aa ntec position sensitive detector (PSD, 4 degree opening) using Cu K$\alpha_1$ radiation, $\lambda$ = 1.540598 \AA. The measurements were made using a 2$\theta$-range of 20-90\textdegree\ at  296 K and 403 K and a  2$\theta$-range of 35-60\textdegree\ in the temperature range 373 K to 393 K.

\subsection{Neutron powder diffraction}
\label{}
Neutron powder diffraction data were collected on the instrument MEREDIT at the Nuclear Physics Institute in Rez, Czech Republic. The neutron beam was monochromatised by a copper mosaic monochromator (reflection 220) giving a wavelength of $\lambda$ = 1.46 \AA. Samples were studied in a 2$\theta$-range of 4-148\textdegree\ at 296 K and 450 K.

\subsection{Refinements of the crystal and magnetic structure}
Structure refinements were performed on the neutron powder profiles by the Rietveld method \cite{rietveld} using the software \textsc{fullprof} \cite{fullprof} and unit cell parameters from XRD data were refined using the software \textsc{unitcell} \cite{unitcell}. The neutron wavelength was refined by using unit cell parameters determined from XRD data as starting point. The refined wavelength was used to refine the unit cell parameters at 450 K from neutron powder diffraction data. The peak shape was described by a psuedo-Voight profile function and the background was determined by a linear interpolation between chosen points. The following parameters in the 450 K data set were varied: peak shape, unit cell parameters, scale factor, half width parameters, zero point, background, atom occupancies, isotropic temperature parameter and atomic coordinates. The same parameters of the 296 K data set were varied  as well as the parameters for the magnetic moments. The magnetic form factors of Fe and Mn were set as shown in Refs. \cite{fuji-79, takei-63}, respectively. The occupancies of the P/Si sites were kept at a 50/50 ratio, thus not allowing the actual P and Si content to be determined from these refinements.
% The corresponding, and unknown, factor for the Mn atoms was varied between 0, +2 and +4 to find the best fitting.

\section{Results}
\label{results}

\subsection{Phase analysis and crystal structure}
\label{structure}
\begin{figure}[t!]
\centering
\includegraphics[width = 75mm]{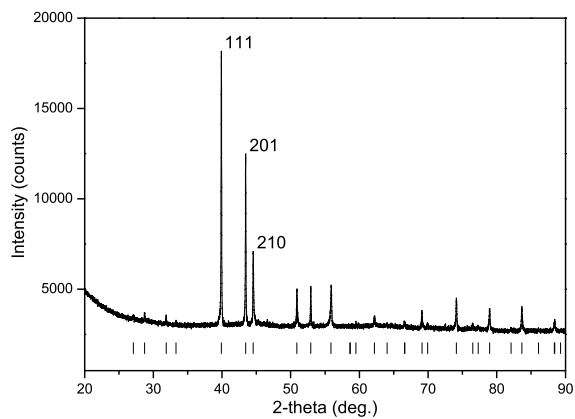}
\caption{X-ray powder diffraction pattern of FeMnP$_{0.5}$Si$_{0.5}$ at 296 K. The tick marks indicate the Bragg positions of FeMnP$_{0.5}$Si$_{0.5}$. $\lambda$ = 1.540598 \AA.}
\label{xray1}
\end{figure}

\begin{figure}[b!]
  \centering
\includegraphics[width = 60mm]{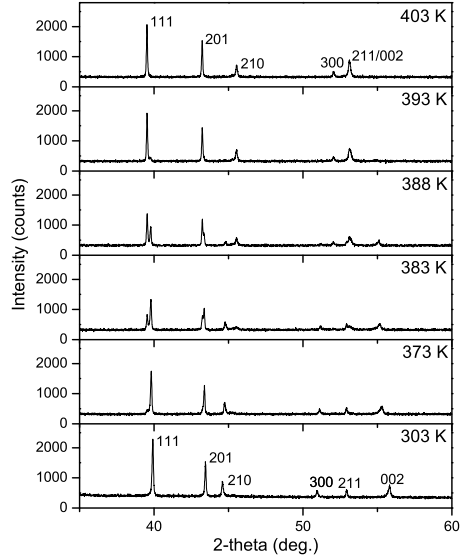}
\caption{XRD-patterns of FeMnP$_{0.5}$Si$_{0.5}$  showing the structural transition occurring at approx. 385 K. $\lambda$ = 1.540598 \AA.}
\label{xray_thermo}
\end{figure}

 The XRD investigation confirms that FeMnP$_{0.5}$Si$_{0.5}$ crystallizes in the hexagonal Fe$_2$P-type structure, space group P\=62m and unit cell parameters a=6.2090(3) \AA, c=3.2880(3) \AA. The XRD pattern for FeMnP$_{0.5}$Si$_{0.5}$ at 296 K is shown in Figure \ref{xray1} which reveals a pattern of a single phase sample.  XRD-patterns in the range 363 to 403 K are shown in Figure \ref{xray_thermo} where it can be seen that FeMnP$_{0.5}$Si$_{0.5}$  undergoes a structural transition (within the space group) between $\sim$373 K and 393 K. The a-axis has decreased $\sim$2\% while the c-axis has increased $\sim$5\% compared to 296 K, see Table \ref{table:fe2p}. The factor c/a and the volume have increased $\sim$7\% and 1\% respectively. The structural transition occurs in the same region as the Curie temperature why it is likely that the transition originate from magnetostriction effects. Structure refinements of powder neutron diffraction data shows that the Fe and Mn atoms are preferably situated in the 3g and 3f sites respectively which are based on the interatomic distances (see Table \ref{table:inter-atom-dist}) and the refined occupancies of the Fe and Mn atoms (see Table \ref{table:occupancy}).

The composition based on the refined occupancies extracted from the neutron powder diffraction data  (however assuming fixed ratio P/Si=1) indicates that the acquired composition of the metallic atoms in the sample is (Fe$_{1.018(3)}$Mn$_{0.982(3)}$P$_{0.5}$Si$_{0.5}$. Also, the synthetic process with stoichiometric amounts of the raw materials showed minor (less than 0.5\%) losses.

%and was affirmed to be a single phase sample by means of X-ray and neutron powder diffraction, see Figure \ref{xray1} and \ref{neutron_diff}, why it is likely that the reported sample is FeMnP$_{0.5}$Si$_{0.5}$.

\begin{table}[b!]
\centering
\caption{Unit cell parameters of  FeMnP$_{0.5}$Si$_{0.5}$ at 296 K and 450 K refined from XRD data, $\lambda$ = 1.540598 \AA.}
% and 450 K (refined from neutron powder diffraction data, Refined wavelength $\lambda$ = 1.4595 \AA).
\begin{tabular}{ l c c c c}
\toprule
T (K) & a (\AA)  & c (\AA) & V (\AA$^3$) & c/a \\
\midrule
296 & 6.2090(3)  & 3.2880(2) & 109.78(2)  & 0.5296(1) \\
450 & 6.0830(8)  & 3.4507(9) & 110.58(4)  & 0.5672(1) \\
\bottomrule
\label{table:fe2p}
\end{tabular}
\end{table}

\begin{table}[b!]
\centering
\caption{Interatomic distances in  FeMnP$_{0.5}$Si$_{0.5}$ at 296 K.}
\begin{tabular}{l l l l l l}
\toprule
Atoms & & Distance (\AA) & Atoms & & Distance (\AA) \\
\midrule
Fe(1) -- & 2 P/Si(2)  & 2.292(2) & Mn(2) -- & 1 P/Si(2)  & 2.502(6) \\
&  2 P/Si(1)&  2.342(2) &  & 4 P/Si(1)  & 2.506(2) \\
& 2 Mn(2)  & 2.675(5) &  & 2 Fe(1)  & 2.675(5) \\
& 4 Mn(2)  & 2.742(1) &  & 4 Fe(1)  & 2.742(4) \\
& 2 Fe(1)  & 2.766(3) &  & 4 Mn(2)  & 3.275(6) \\
\\
P/Si(1) --  & 3 Fe(1)  & 2.342(2) & P/Si(2)  -- & 6 Fe(1)  & 2.292(2) \\
& 6 Mn(2)  & 2.506(5) &  & 3 Mn(2)  & 2.502(3) \\
\bottomrule
\label{table:inter-atom-dist}
\end{tabular}
\end{table}

\begin{figure}[t]
\centering
\includegraphics[width = 70mm]{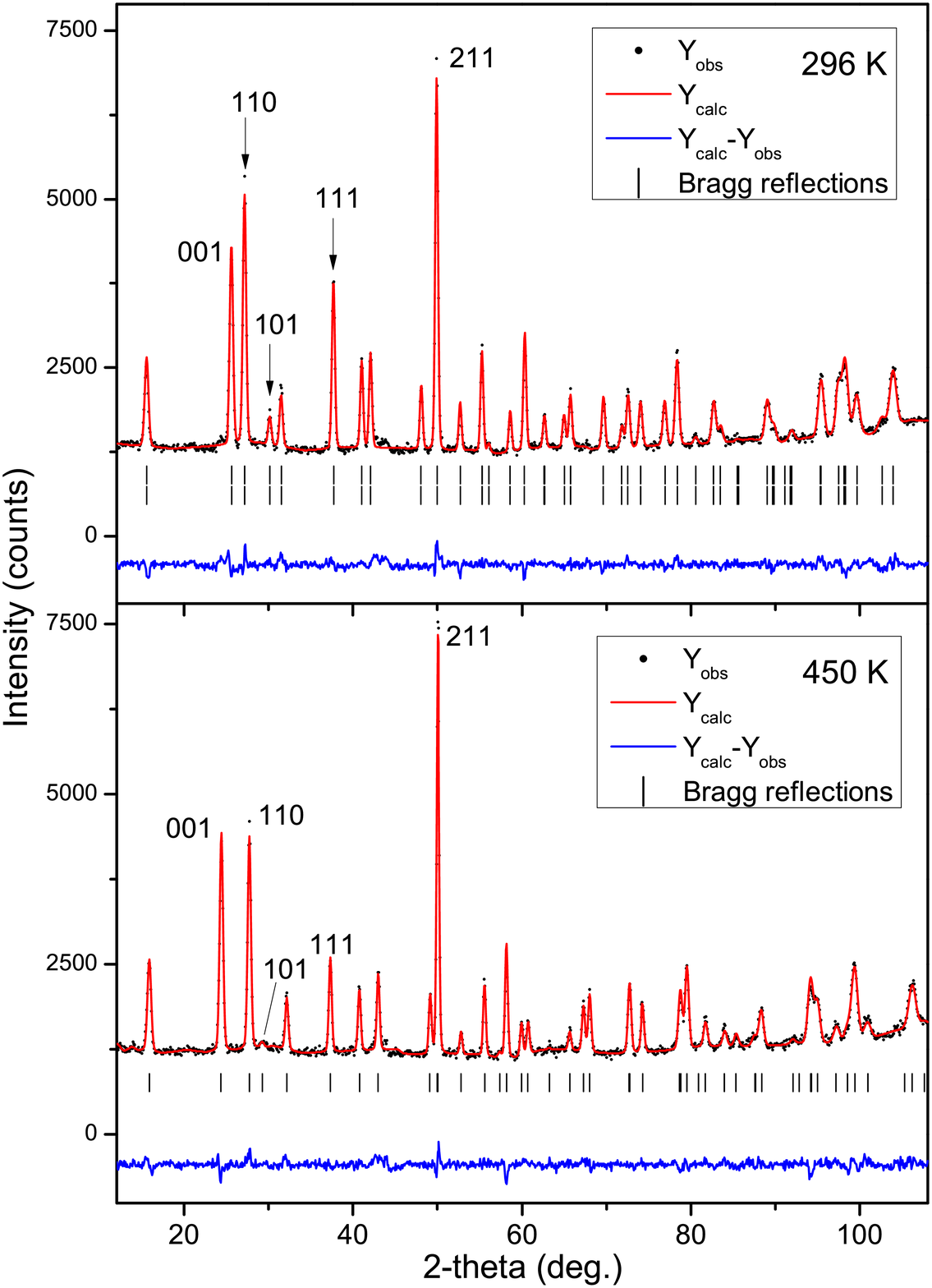}
\caption{Structure refinements from neutron powder diffraction data of  FeMnP$_{0.5}$Si$_{0.5}$ at 296 K and 450 K. The differences in intensity and position of the Bragg peaks are due to the ferro- to paramagnetic transition at $\sim$390 K and the structural transition at $\sim$385 K. The peaks with the highest magnetic intensity are marked with an arrow.}
\label{neutron_diff}
\end{figure}

\subsection{Magnetic structure}
Refinements of the neutron powder diffraction data at 296 K and 450 K are shown in Figure \ref{neutron_diff}. The magnetic contribution from the ferromagnetic phase at 296 K is distinguished by the higher peak-intensities at lower 2-theta angles. The magnetic moments are aligned in the a-direction and are 1.9(1) $\mu_B$ and 2.5(1) $\mu_B$ in the M(1) and M(2) site respectively which gives a total magnetic moment of 4.4(2) $\mu_B$. The magnetic symmetry group was found to be Cm2m.

%The ferromagnetic behavior below T$_C$ 398 K can  The differences in peak-intensity originates from the magnetic contribution of the ferromagnetic state below  T$_C$. The magnetic contributions at 296 K was extracted from structure refinements.

% The occupancy and positions of the Mn and Fe atoms are shown in Table \ref{table:occupancy}.

%The Fe atoms prefer the tetrahedral 3g site while the Mn atoms are mainly situated in the pyramidal 3f site.

%A summary of the different magnetic form factors can be seen in Table \ref{table:mag-form-factor} together with the magnetic moment of the two metal sites.   The magnetic form factor of the Mn(1) and Mn(2) atoms was set to 0 and +4 respectively which resulted in magnetic moment of 2.0(1) $\mu_B$ and 2.2(1) $\mu_B$ in the M(1) and M(2) site respectively.

\begin{table*}[t!]
  \centering
  \caption{Placement and occupancy of the Fe and Mn atoms in FeMnP$_{0.5}$Si$_{0.5}$ at 296 K and 450 K. Derived from refinements of neutron powder diffraction data. (0, 0, 0) was chosen as an origin. }
\begin{tabular}{lllllllllllll}
\toprule
 &  &  & 296 K & & & & & 450 K & & &\\
Atom & Site & & x & y & z &  Occ. &  & x & y & z &  Occ.\\
\midrule
Fe(1) & 3g & & 0.2572(4) & 0 & 1/2  & 0.2353(4) & & 0.2548(3) & 0 & 1/2   & 0.2353(4)\\
Mn(1) & 3g & & 0.2572(4) & 0 & 1/2  & 0.0147(4) & & 0.2548(3) & 0 & 1/2  & 0.0147(4)\\
Fe(2) & 3f & & 0.597(1) & 0 & 0  & 0.0193(2) & & 0.591(1) & 0 & 0  & 0.0193(2) \\
Mn(2) & 3f & & 0.597(1) & 0 & 0  &  0.2307(2) & & 0.591(1) & 0 & 0  & 0.2307(2) \\
P/Si(1) & 2d & & 1/3 & 2/3 & 1/2  & 0.1667 & & 1/3 & 2/3 & 1/2 & 0.1667\\
P/Si(2) & 1a & & 0 & 0 & 0  & 0.0833 & & 0 & 0 & 0 &  0.0833 \\
\midrule
 & & & \multicolumn{4}{l}{R$_p$ = 2.37\%, R$_{wp}$ = 3.05\%, $\chi^2$ = 3.03} & & \multicolumn{4}{l}{R$_p$ = 2.59\%, R$_{wp}$ = 3.36\%, $\chi^2$ = 3.44} & \\
 & & & \multicolumn{4}{l}{R$_{Bragg}$ = 5.26\%, R$_{mag}$ = 6.08\% }  & & \multicolumn{4}{l}{R$_{Bragg}$ = 4.53\%} & \\
\bottomrule
\label{table:occupancy}
\end{tabular}
\end{table*}

%\begin{table}[t!]
%  \centering
%\caption{Different sets of magnetic form factors for the Mn atoms and the resulting magnetic moment in the two Mn positions at 296 K. The corresponding factor for the Fe(1) and Fe(2) atoms was set to 0 and +4 respectively. Row in bold was used in the final refinements.}
%\begin{tabular}{c c c c c c c}
%\toprule
%\multicolumn{4}{c}{Magn. form factors} & \multicolumn{3}{c}{Magnetic moments ($\mu_B$)}\\
%& Mn(1) & Mn(2) & & M$_1$ & M$_2$ & M$_{tot}$ \\
%\midrule
%& 0 & 0 & & 2.0(1) & 2.7(2) & 4.7(3)\\
%& +2 & 0 & & 2.0(1) & 2.7(2) & 4.7(3)\\
%& +4 & 0 & & 2.0(1) & 2.7(2) & 4.7(3)\\
%& 0 & +2 & & 2.0(1) & 2.6(1) & 4.6(3)\\
%& +2 & +2 & & 2.0(1) & 2.6(2) & 4.6(3)\\
%& +4 & +2 & & 1.9(1) & 2.6(2) & 4.5(3)\\
%& \textbf{0} & \textbf{+4} & & \textbf{1.9(1)} & \textbf{2.3(1)} & \textbf{4.2(2)}\\
%& +2 & +4 & & 1.9(1) & 2.3(1) & 4.2(2)\\
%& +4 & +4 & & 1.9(1) & 2.3(1) & 4.2(2)\\
%\bottomrule
%\label{table:mag-form-factor}
%\end{tabular}
%\end{table}

\section{Discussion}
\label{}

\begin{figure}[b!]
\centering
\includegraphics[width = 80mm]{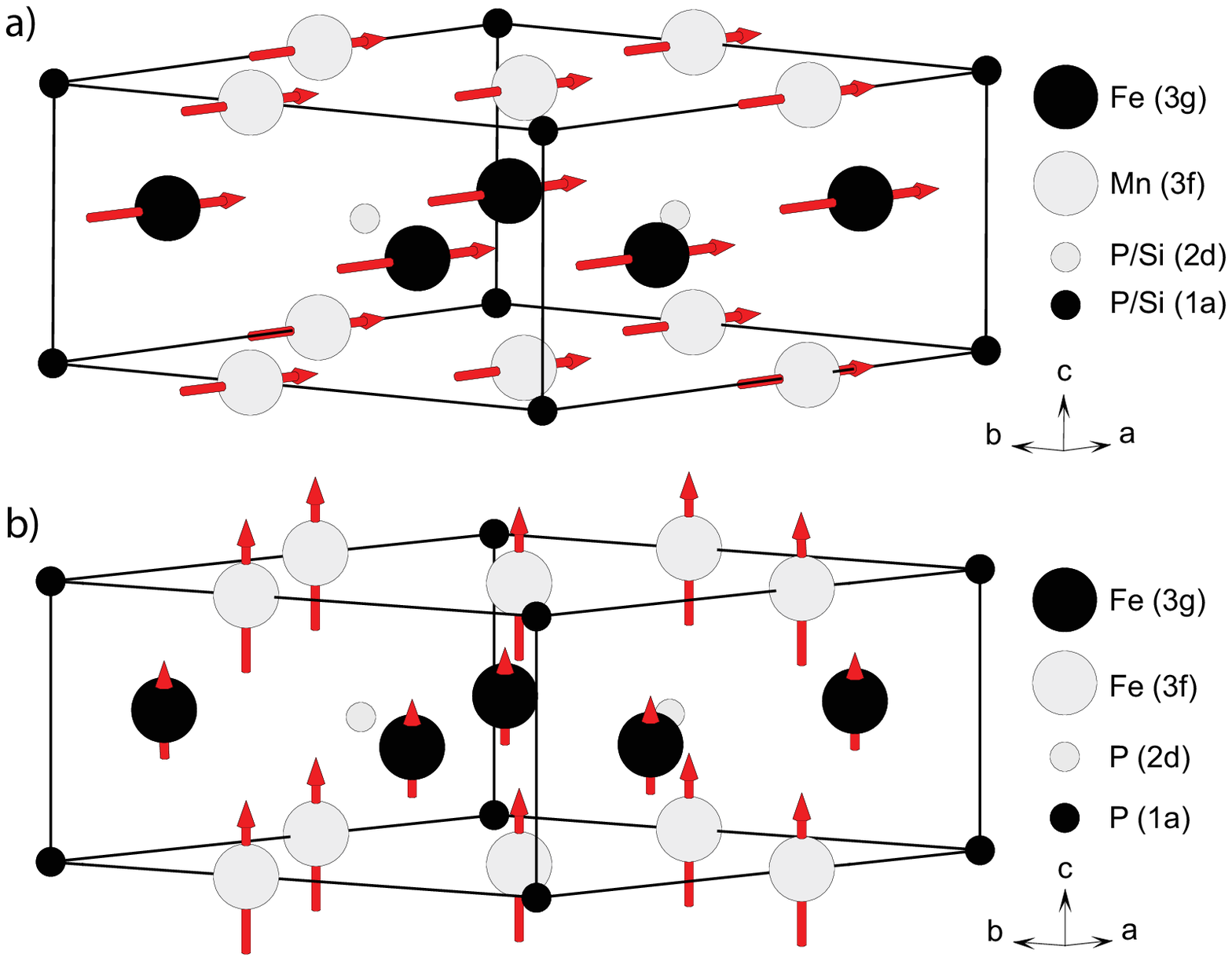}
\caption{The magnetic structure of FeMnP$_{0.5}$Si$_{0.5}$, a), and Fe$_2$P, b). The magnetic moments in Fe$_2$P are aligned in the c-direction while the moments in FeMnP$_{0.5}$Si$_{0.5}$ are aligned in the a-direction. The length of the arrows correspond to the magnitude of the magnetic moments.}
\label{magn_str}
\end{figure}

%The present study was performed on a single phase sample of FeMnP$_{0.5}$Si$_{0.5}$ produced by a careful and well controlled synthetic method. The composition based on the refined occupancies extracted from the neutron powder diffraction data indicates a possible minor deviation of the composition (Fe$_{1.018(3)}$Mn$_{0.982(3)}$P$_{0.5}$Si$_{0.5}$. However, the synthetic process with stoichiometric amounts of the raw materials showed minor (less than 0.5\%) losses and was affirmed to be a single phase sample by means of X-ray and neutron powder diffraction, see Figure \ref{xray1} and \ref{neutron_diff}, why it is likely that the reported sample is FeMnP$_{0.5}$Si$_{0.5}$.

The magnitude of the magnetic moments of the Fe atoms and the total magnetic moment per formula unit is confirmed by recent  M\" ossbauer and magnetization studies of  FeMnP$_{0.5}$Si$_{0.5}$ \cite{mossbauer_uu}. The total magnetic moment has been reported to be approx. 4.4 $\mu_B$/f.u which is in good agreement with our result, where the total magnetic moment is 4.4(2) $\mu_B$/f.u. The size of the total magnetic moment is higher than the corresponding value for Fe$_{2}$P (2.9(1) $\mu_B$) \cite{fujii-87, koumina, scheerlinck, tobola}. Earlier studies have also shown that there are significant differences in magnitude of the moments between the two Fe sites in  Fe$_{2}$P. The Fe atom in the pyramidal 3f site possesses a smaller magnetic moment than the Fe atom in the tetrahedral 3g site, which is illustrated in Figure \ref{magn_str}b.

The refinements of the neutron powder diffraction intensities of FeMnP$_{0.5}$Si$_{0.5}$ indicate that the Mn atoms prefer to be situated in the pyramidal 3f site in the Fe$_{2}$P-structure. The substitution of Fe with Mn in the 3f site increases the magnetic moment in the site with $\sim$0.8 $\mu_B$  compared to Fe$_{2}$P. The magnetic moment of the Fe atom in the 3g site of FeMnP$_{0.5}$Si$_{0.5}$ is also shown to increase $\sim$0.6 $\mu_B$.

The magnetic structures of FeMnP$_{0.5}$Si$_{0.5}$ and Fe$_{2}$P are shown in Figure \ref{magn_str}. The magnetic moments are aligned in the a-direction while the moments of Fe$_{2}$P are aligned in the c-direction \cite{lundgren_fe2p, fujii_c_axis}. A similar alignment of the magnetic moments as in FeMnP$_{0.5}$Si$_{0.5}$ has also been reported to occur in FeMnP$_{0.5}$As$_{0.5}$ \cite{bacmann-94}. It is of interest to note that in FeMnP$_{0.7}$As$_{0.3}$ the moments deviate from the c-axis by 50\textdegree.

% where FeMnP$_{0.5}$As$_{0.5}$ shows a similar alignment of the magnetic moments as in FeMnP$_{0.5}$Si$_{0.5}$. Refinements of neutron powder diffraction data of FeMnP$_{0.7}$As$_{0.3}$ have shown that the magnetic moments deviate from the c-axis by 50\textdegree. It is likely that corresponding alloying in the  FeMnP$_{1-x}$Si$_{x}$-system will show comparable results.

Magnetoelastic transitions have been found for the hexagonal system FeMnP$_{1-y}$As$_{y}$ (0.15 $\le$ y $\le$ 0.66) isostructural with FeMnP$_{0.5}$Si$_{0.5}$. A marked increase both in magnetic hyperfine field and in magnetic moment take place on the Fe tetrahedral site (Fe$_t$) as compared to the tetrahedral site in Fe$_{2}$P \cite{bacmann-94, malaman-1996}. Table \ref{table:mag-state} displays the interatomic distances in paramagnetic and ferromagnetic Fe$_{2}$P, FeMnP$_{0.7}$As$_{0.3}$ and the presently studied compound FeMnP$_{0.5}$Si$_{0.5}$.

\begin{table}[t!]
  \centering
\caption{Magnetic state, Fe saturation magnetic moment and interatomic average distances for tetrahedral Fe$_t$ in FeMX with Fe$_2$P structure, M$_p$ = pyramidal Fe or Mn and X = P, Si and As.}
\begin{tabular}{ l c c c c c c c}
\toprule
 & Magn.  &  Fe$_t$ sat.  &  &  Fe$_t$-X &  Fe$_t$- Fe$_t$ &  Fe$_t$- M$_p$  &  \\
Compound & state (T$_C$)   & mom. ($\mu_B$) & V (\AA$^3$) & (\AA) & (\AA) & (\AA)  & Ref. \\
\midrule
Fe$_{2}$P & PM (295 K)  & - & 103.1  & 2.255 & 2.610 & 2.682 & \cite{hf_ugn} \\
 & FM (77 K) & 1.03 & 102.9 & 2.253 & 2.597 & 2.682 & \cite{scheerlinck}\\
FeMnP$_{0.7}$As$_{0.3}$ & PM (250 K)  & - & 110.7  & 2.311 & 2.638 & 2.755 & \cite{bacmann-94} \\
 & FM (100 K) & 1.25 & 110.2 & 2.312 & 2.747 & 2.743 & \cite{bacmann-94}\\
FeMnP$_{0.5}$Si$_{0.5}$ & PM (450 K) & - & 109.8 & 2.312 & 2.685 & 2.726 & This work\\
 & FM (296 K) & 1.65 & 110.6 & 2.317 & 2.766 & 2.708 & This work\\
\bottomrule
\label{table:mag-state}
\end{tabular}
\end{table}

As can be seen from the table the cell volume, the average near Fe$_t$--X distances and Fe$_t$--M$_p$ do not change significantly between the paramagnetic (PM) and the ferromagnetic (FM) state. However the crystal a- and c-axis decreases and increases, respectively, when passing the first order ferromagnetic transition from lower temperature. A large interatomic difference is however observed for the Fe$_t$--Fe$_t$ distances for the  high moment cases of FeMnP$_{0.7}$As$_{0.3}$ and FeMnP$_{0.5}$Si$_{0.5}$. The cell volume expansion between the three different compounds is obvious. The overall crystal expansion and the increase in Fe$_t$--Fe$_t$  distances yields a stronger Fe$_t$  electron localization and results in a larger Fe$_t$  magnetic moment.

\section{Conclusions}
\label{conclusions}

The magnetocaloric compound FeMnP$_{0.5}$Si$_{0.5}$ has been synthesized and studied regarding the crystallographic and magnetic structure. X-ray and neutron powder diffraction experiments show that the sample is single phase and reveals the magnetic structure of FeMnP$_{0.5}$Si$_{0.5}$. An isotructural phase transition has been observed at about the same temperature as the magnetic phase transition (382 K). The Fe atoms are mainly situated in the tetrahedral 3g site while the Mn atoms prefer the pyramidal 3f position. The magnetic moments derived from neutron powder diffraction are shown to be coordinated along the a-axis with a total moment of 4.4 $\mu_B$. This high value of the magnetic moment goes along with our M\"ossbauer results \cite{mossbauer_uu} and is in accord with a strong increase in the Fe$_t$-Fe$_t$ distances. The high magnetic moment and a readily tunable transition temperature make (slightly) off-stoichiometric FeMnP$_{0.5}$Si$_{0.5}$ a promising alloy system for magnetocaloric applications.

\section*{Acknowledgments}
This work was financed by the Swedish Research Council and the Swedish Energy Agency, which is gratefully acknowledged.



%% References


\bibliographystyle{model1a-num-names}
\bibliography{bibtexreferences}






% \end{thebibliography}

\end{document}